\documentclass[pra,twocolumn,superscriptaddress,showpacs,amsmath,amstex,amssymb,citeautoscript]{revtex4-1}

\usepackage[compat=1.1.0]{tikz-feynman}
\usepackage[T1]{fontenc}
\usepackage[utf8]{inputenc}
\usepackage{lipsum}
\usepackage{amsmath}
\usepackage{amssymb}
\usepackage{bbm}
\usepackage{braket}
\usepackage{xcolor}
\usepackage{pifont}
\usepackage[mathscr]{euscript}
\usepackage[shortlabels]{enumitem}
\usepackage[justification=justified]{subcaption}
\usepackage{graphicx}
\usepackage{lipsum}
\allowdisplaybreaks
\usepackage{float}
\usepackage{graphicx}
\usepackage{dsfont}
\usepackage{comment}
\usepackage[colorlinks=true]{hyperref}  
\hypersetup{
    bookmarks=true,         
    unicode=false,          
    pdftoolbar=true,        
    pdfmenubar=true,        
    pdffitwindow=false,     
    pdfstartview={FitH},    
    pdftitle={TITLE},    
    pdfauthor={Someone et al.},     
    pdfsubject={},   
    pdfcreator={},   
    pdfproducer={}, 
    pdfkeywords={} {} {}, 
    pdfnewwindow=true,      
    colorlinks=true,       
    linkcolor=blue, 
    citecolor=blue,        
    filecolor=magenta,      
    urlcolor=blue           
}

\newcommand{\equref}[1]{Eq.~(\ref{#1})}
\newcommand{\equsref}[2]{Eqs.~(\ref{#1}) and (\ref{#2})}

\newcommand{\figref}[1]{Fig.~\ref{#1}}
\newcommand{\refcite}[1]{Ref.~\onlinecite{#1}}

\newcommand{\appref}[1]{Appendix~\ref{#1}}

\newcommand{\pdagger}{{\phantom{\dagger}}}

\renewcommand{\vec}[1]{\boldsymbol{#1}}

\definecolor{wrongultramarine}{rgb}{1,0.5,0}

\begin{document}

\title{Dissipation-enhanced non-reciprocal superconductivity: \\ application to multi-valley superconductors}

\author{Sayan Banerjee}
\affiliation{Institute for Theoretical Physics III, University of Stuttgart, 70550 Stuttgart, Germany}

\author{Mathias S.~Scheurer}
\affiliation{Institute for Theoretical Physics III, University of Stuttgart, 70550 Stuttgart, Germany}

\begin{abstract}
We here propose and study theoretically a non-equilibrium mechanism for the superconducting diode effect, which applies specifically to the case where time-reversal-symmetry---a prerequisite for the diode effect---is spontaneously broken by the superconducting electrons themselves. 
We employ a generalized time-dependent Ginzburg-Landau formalism to capture dissipation effects in the non-equilibrium current-carrying state via phase slips and show that the coupling of the resistive current to the symmetry-breaking order is enough to induce a diode effect. Depending on parameters, the critical current asymmetry can be sizeable, asymptotically reaching a perfect diode efficiency; the competition of symmetry-breaking order, superconducting and resistive currents gives rise to rich physics, such as current-stabilized, non-equilibrium superconducting correlations. Although our mechanism is more general, the findings are particularly relevant to twisted trilayer and rhombohedral tetralayer graphene, where the symmetry-breaking order parameter refers to the imbalance of the two valleys of the systems. 
\end{abstract}

\maketitle
The superconducting diode effect (SDE) has been a subject of major interest recently \cite{Review} due to its potential applications to superconducting electronics and its relation to the competition between superconductivity and applied external fields or coexisting time-reversal breaking orders. 
The SDE requires broken time-reversal and inversion symmetries and is characterized by different critical supercurrent densities in opposite directions $\hat{n}$ and $-\hat{n}$, i.e., $j_{c}(\hat{n}) \neq j_{c}(-\hat{n})$. Various systems have been concocted to show the SDE \cite{ando_observation_2020,lyu_superconducting_2021,du_superconducting_2023,sundaresh_diamagnetic_2023,kealhofer_anomalous_2023,hou_ubiquitous_2023,chen_superconducting_2023,gupta_superconducting_2022,AnotherJosephson,banerjee_phase_2023,pal_josephson_2022,kim_intrinsic_2023,gupta_superconducting_2022,MoreJosephson,ParadisoNbSe2,wu_field-free_2022,diez-merida_magnetic_2021,golod_demonstration_2022,chiles_non-reciprocal_2022-1,zhang_reconfigurable_2023,shin_magnetic_2021,jeon_zero-field_2022,kealhofer_anomalous_2023,hou_ubiquitous_2023,lin_zero-field_2022,anwar_spontaneous_2022,narita_field-free_2022,gutfreund_direct_2023} and a variety of theoretical works \cite{daido_superconducting_2022,daido_intrinsic_2022,yuan_supercurrent_2022,he_phenomenological_2022,ilic_theory_2022,scammell_theory_2022,zinkl_symmetry_2022,PhysRevX.12.041013,he_supercurrent_2022,PhysRevB.106.L140505,jiang_field-free_2022,kokkeler_field-free_2022,chazono_piezoelectric_2022,vodolazov_superconducting_2005,de_picoli_superconducting_2023,kochan_phenomenological_2023,ikeda_intrinsic_2022,tanaka_theory_2022,wang_symmetry_2022,haenel_superconducting_2022,legg_parity_2023,cuozzo_microwave-tunable_2023,souto_josephson_2022,cheng_josephson_2023,steiner_diode_2022,costa_microscopic_2023,wei_supercurrent_2022,legg_superconducting_2022,karabassov_hybrid_2022,2022arXiv221114846H,PhysRevLett.130.126001,guo__0-junction_2024,banerjee_altermagnetic_2024,karabassov_competitive_2024,hosur_digital_2024,cayao_enhancing_2024,debnath_gate-tunable_2024,cadorim_harnessing_2024-1,kokkeler_nonreciprocal_2024,bhowmik_optimizing_2024,chakraborty_perfect_2024,angehrn_relations_2024,shaffer_superconducting_2024,huang_superconducting_2024,roig_superconducting_2024,hasan_supercurrent_2024,seoane_souto_tuning_2024,daido_unidirectional_2023, sim2024pairdensitywavessupercurrent} have accompanied the experiments, in order to explain the asymmetry in the critical current of the respective superconductor. 

Of particular note is an experiment with twisted trilayer graphene \cite{lin_zero-field_2022}, where a pronounced asymmetry has been observed in the critical current in the absence of external magnetic fields. The intrinsic nature of such a SDE arising from interaction-induced time-reversal symmetry (TRS) breaking orders was explored theoretically in \cite{scammell_theory_2022} and shown to be most naturally consistent with an imbalance in the population of the system's two valleys---we refer to the associated order parameter as valley polarization (VP) in the following. Furthermore, an extreme asymmetry was observed, quantified by an efficiency parameter $\eta(\hat{n}) = \frac{|j_c(\hat{n})-j_c(-\hat{n})|}{j_c(\hat{n}) + j_c(-\hat{n})}$ close to $1$, with the critical current nearly vanishing in one direction while remaining non-zero in the other. A follow-up study \cite{our_backaction} which considered a back-action mechanism in which the non-dissipative supercurrent couples back to VP showed results of an enhanced diode effect. Even though this theory could explain diode efficiencies up to 60-65 \%, it is unable to capture the extreme asymmetry observed in the experiment. 

Meanwhile, it was shown in \cite{serlin_intrinsic_2020,sharpe_emergent_2019} that \textit{normal}, dissipative currents can induce sign reversals of Hall conductance and also flip the VP; \refcite{balents_current} developed a theory for how  VP can couple to the normal current. 
These findings suggest the need to employ a more general time-dependent framework to take into account dissipation effects and resistive currents in the description of the SDE in systems where time-reversal symmetry is spontaneously broken by the electron liquid. Also beyond that, there are always phase slips in the bulk of the superconductors as one increases the current flow. A true diode effect calculation would therefore be remiss without taking such non-equilibrium effects into account.

Motivated by these ideas, in this letter, we investigate zero-field non-reciprocal superconductivity within a generalized time-dependent Ginzburg-Landau (GL) framework where we incorporate a coupling of the dissipative current in the system to the underlying TRS-breaking order parameter. We show that this coupling alone is sufficient to induce a SDE, providing a mechanism for the SDE that is completely distinct from the frequently studied field- or order-parameter-induced TRS-breaking momentum dependencies of the particle-particle bubble or Cooper-pair susceptibility, see, e.g., \cite{daido_superconducting_2022,daido_intrinsic_2022,yuan_supercurrent_2022,scammell_theory_2022}. On top of that, we analyze different parameter regimes, which illustrate the complex interplay of superconductivity, the symmetry-breaking normal-state order, and dissipation. This further reveals that this mechanism allows for, in principle, arbitrarily large current asymmetries with $\eta \rightarrow 1$. 
We note that our study is different from \refcite{daido_unidirectional_2023} where the SDE was studied along a direction perpendicular to a fixed additional dissipative current. Finally, apart from twisted trilayer  \cite{lin_zero-field_2022}, our mechanism is also likely relevant for rhombohedral graphene where recent experiments \cite{rhombohedral} have found signs of VP to coexist with superconductivity and a SDE is expected \cite{scammell_theory_2022}.

\textit{Time-dependent GL formalism.---}As motivated above, we here study the SDE including dissipative currents using the generalized time-dependent GL equations. The equation governing the dynamics and spatial profile of the complex superconducting order parameter $\psi(\vec{x},t)$ reads in dimensionless units as \cite{TDGL_numerics} 
\begin{align}
\begin{split}
\frac{u}{\sqrt{1+\gamma^2|\psi|^2}} &\left(\frac{\partial}{\partial t}+i \mu+\frac{\gamma^2}{2} \frac{\partial|\psi|^2}{\partial t}\right) \psi \\ &=\left(\epsilon-|\psi|^2\right) \psi +(\vec{\nabla}-i \vec{A})^2 \psi.
    \label{full_tdgl}
    \end{split}
\end{align}
Here $\mu$ is the electric scalar potential, $\epsilon$, which can be computed as a one-loop diagram from microscopic theory (see \appref{appendix:particlebubble}), determines the strength of superconductivity, and $u$ is related to the ratio of the relaxation times for amplitude and phase of the order parameter in dirty superconductors. Furthermore, the parameter $\gamma$ characterizes the strength of inelastic electron-phonon scattering.  The total current $\vec{j}$ is the sum of supercurrent $\vec{j}_{s} =  \text{Im}[\psi^{*} (\nabla- ie\vec{A})\psi]$ and the normal current contribution $\vec{j}_{N}= - \vec{\nabla} \mu$. This set of equations is supplemented with the electroneutrality condition $\vec{\nabla}\cdot\vec{j} = 0$.

\begin{figure}[tb]
    \centering
\includegraphics[width=\linewidth]{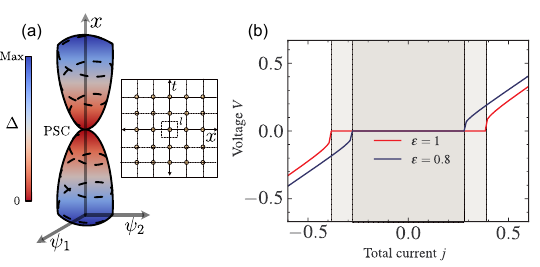}
    \caption{ (a) Dynamics of the order parameter at an instant in time (left) where a PSC occurs and on the right PSCs (denoted by circles) in space-time; (b) Current-Voltage characteristics for $\epsilon = 1$ ($\Phi_V=0$) and $\epsilon=0.8$ ($\Phi_V \neq 0$). The gray shadings denote the region with non-dissipative superconductivity, i.e.~no normal current, $j_N=0$, and zero voltage drop, $V=0$.}
    \label{fig:VI_characteristics_2}
\end{figure}

The first step to capture a dissipative current carrying state lies in understanding the phenomenon of resistive states in superconductors \cite{resistive1,psc_new,resistive_ref,ivlev_dynamic_1980,ivlev_dynamical_1984,ivlev_dynamics_1983}. One way to describe these resistive states is to consider the theory of phase slips \cite{psc_old,skocpol_phase-slip_1974} in (quasi) one-dimensional systems with the following basic physical picture. Applying an electric field results in the acceleration of Cooper pairs, which naively would give rise to a phase difference between two distant points in the superconductor that increases with time. This would mean that the ``helix'' shown as a dashed line in \figref{fig:VI_characteristics_2}(a)(left), where we split the order parameter into real and imaginary parts, $\psi = \psi_{1} + i\psi_{2}$, becomes increasingly dense. To avoid this, the system develops phase slip centers (PSC) where the modulus $\Delta = \sqrt{\psi_{1}^{2} + \psi_{2}^{2}}$ of the order parameter locally drops to zero, as indicated in \figref{fig:VI_characteristics_2}(a)(left), allowing to ``shake off'' a phase difference which is a multiple of $2\pi$.
For superconductivity to coexist with a finite electric field in the bulk, we thus need a finite density (in space) of PSCs appearing at a finite rate.
In \figref{fig:VI_characteristics_2}(a)(right), we illustrate the periodic occurrence of PSCs (denoted by circles) in space-time. Traversing the indicated loop $l$ results in the phase of the order parameter changing by a multiple of $2\pi$, leading the PSCs to behave as topological singularities in two-dimensional space-time \cite{resistive_ref}.

To quantify this phenomenon further, we closely follow \cite{resistive_ref,ivlev_theory_1984} and consider the case of a one-dimensional system in the absence of an external magnetic field (we choose $\vec{A}=0$) in the limit $\gamma \gg 1$. Taking advantage of the fact that the region with significant voltage variations around the PSC is much larger than the dynamical region very close to the PSCs, where both the order parameter and the current are oscillatory, one can derive the equations (see \appref{appendix:resistive_state} for more details and complete derivation as well as \cite{resistive_ref})
\begin{subequations}
\begin{equation}
    V \int_{\Delta_0}^{\sqrt{\epsilon}} \frac{\left(3 \Delta^2-2\epsilon\right) d \Delta}{\sqrt{\left(\epsilon-\Delta^2\right)} F_{\Phi}\left(\Delta, \Delta_0\right)}=2 F_{\Phi}\left(\sqrt{\epsilon}, \Delta_0\right)\Theta(j-j_{s0})
\end{equation}
\begin{equation}
    F_{\Phi}^2\left(\Delta, \Delta_0\right)=\int_{\Delta_0}^{\Delta} \frac{\left[j-x^2\sqrt{\left(\epsilon-x^2\right)}\right]\left(3 x^2-2\epsilon\right) d x}{\sqrt{\left(\epsilon-x^2\right)}},
\end{equation}\label{eq:VI_characteristics}\end{subequations}
from time averages of the time-dependent GL equations. We will use these expressions to solve for the voltage drop $V$ for a given current $j$.
In \equref{eq:VI_characteristics}, $\Delta_{0}$ is the value of $\Delta$ midway between two PSCs, $j_{s0}$ the critical supercurrent, $\Theta$ the Heaviside step function, and we assumed $j>0$ for notational simplicity. 
To determine the value of $\Delta_{0}$, we consider the expression for the supercurrent $ j_{\mathrm{s}}=\Delta^2 \sqrt{\left(\epsilon-\Delta^2\right)}$ and maximize it to obtain the critical supercurrent $j_{s0}$
achievable in between the PSCs. 
The critical supercurrent, then given by $j_{s0} = 2 \epsilon^{3/2}/3\sqrt{3}$, fixes the lower bound of the integral $ \Delta_{0} =  \sqrt{2\epsilon/3}$. We also note that if the total applied current is less than $j_{s0}$, the voltage is set to zero and we have a stationary dissipationless superconductor without phase slips inducing voltage drops, i.e.~$j = j_{s}$. This can also be seen by considering the stationary limit of \equref{full_tdgl} with a solution of the form $\psi = \Delta e^{iqx}$. For $j>j_{s0}$, the system transitions to a resistive state. In this regime, we solve the equations \equref{eq:VI_characteristics} above numerically by calculating the voltage $V$ as a function of the total current $j$ with result shown in \figref{fig:VI_characteristics_2}(b). The transition to the Ohmic regime is characterized by the point where the slope of the $V$-$j$ characteristics, or the differential resistance $\frac{dV}{dj}=1$ [recall that we introduced dimensionless units in \equref{full_tdgl}]. Naturally, decreasing $\epsilon$ leads to a destabilization of superconductivity and a reduction of the superconducting region (denoted in gray).

\textit{Valley polarization.---}Equipped with the theory of dissipative superconductivity, we want to explore how a normal-state order parameter breaking time-reversal symmetry, such as VP, defined as a shift in the chemical potentials $\mu_\pm$ in the two valleys, $\Phi_V := (\mu_+-\mu_-)/2$, influences the current-voltage relationship. 
To incorporate the effect of VP in our framework, we parameterize $\epsilon$ in \equref{full_tdgl} to now depend on $\Phi_{V}$ as $\epsilon(\Phi_{V}) = 1 + \beta\Phi_{V}^{2}$ with $\beta<0$, encoding the competitive interplay between VP and superconductivity. Note that time-reversal symmetry prohibits a linear term such that the leading-order contribution is quadratic in $\Phi_V$, see also \appref{appendix:particlebubble}. As $\Phi_V\neq 0$ reduces $\epsilon$, we expect that the critical current is reduced, which we can see in the resulting $V$-$j$ characteristics in \figref{fig:VI_characteristics_2}(b). However, the critical currents are still symmetric and there is no SDE. This is because the $\Phi_V$-induced gradient terms in \equref{full_tdgl}, which break inversion and time-reversal symmetry in the theory, are not yet included. As we here want to focus on a different mechanism for the SDE, we refer to \refcite{scammell_theory_2022} for the study of these terms and neglect them in the following. 

Instead, we will here show how a SDE can result from a coupling of $\Phi_V$ and the dissipative current $j_{N}$. Such a coupling has been demonstrated experimentally \cite{serlin_intrinsic_2020,sharpe_emergent_2019} and can be captured theoretically through a combination of non-equilibrium Keldysh formalism and semi-classical Boltzmann equation, see \cite{balents_current}; in a nutshell, an applied DC current generates a density imbalance in the two valleys which in turn acts as a symmetry-breaking field in the GL equations for $\Phi_V$. This can be compactly reformulated as a GL free-energy,
\begin{equation}
    \mathcal{F}_{\Phi_{V}} = \alpha_{1} \Phi_{V}^{2} + \alpha_{2} \Phi_{V}^{4} + 2 c_{1} j_{N}\Phi_{V}, 
    \label{eq:free_energy}
\end{equation}
to be minimized to find $\Phi_V$. The first two terms determine the value of $\Phi_{V}$ in absence of dissipation, whereas the last term  parameterized by $c_{1}$ captures the aforementioned coupling of $\Phi_{V}$ to a normal state current.
This coupling leads to a back-action mechanism between current and $\Phi_{V}$: 
the VP featuring in  $\epsilon$ determines the normal current $j_{N}$ through \equref{eq:VI_characteristics} which, in turn, determines $\Phi_{V}$ through the minimization of \equref{eq:free_energy}. As this back action will depend on the direction of the current relative to the sign of $\Phi_{V}$, the theory now contains the $\Phi_{V}$-induced symmetry reduction which, as we will see below, gives rise to a SDE. The back action and associated SDE is controlled by the parameter $c_1$ in \equref{eq:free_energy}. 

Concurrently, a coupling ($\propto j_{s}\Phi_{V}$) between the VP and the supercurrent \cite{our_backaction} presents another possible route to enable a similar back-action mechanism in an equilibrium setting. This can also significantly increase the current asymmetry yielding higher values of SDE efficiency $\eta$ compared to the case when the term is absent ($j_{c}^{+} \sim 7 j_{c}^{-}$). However, this approach alone is insufficient to account for and capture the extreme current asymmetries in \refcite{lin_zero-field_2022}. We shall henceforth neglect such a coupling term in the remainder of the manuscript.

In the following, we set $\beta = -1.2$ and redefine $\epsilon = \Theta(\epsilon)\epsilon$. 
This sets the critical VP (at zero current) to be $\Phi_{V}^{c} = 0.91$, since for $\Phi_{V}> \Phi_{V}^{c}$, $\epsilon=0$ which means we are in the normal state, governed by Ohm's law, $V=j$. 
On this basis, in order to explore further the predictions of the theory, we consider three regimes, each with an increasing strength of VP and complexity. 

\begin{figure}[tb]
    \centering
\includegraphics[width=\linewidth]{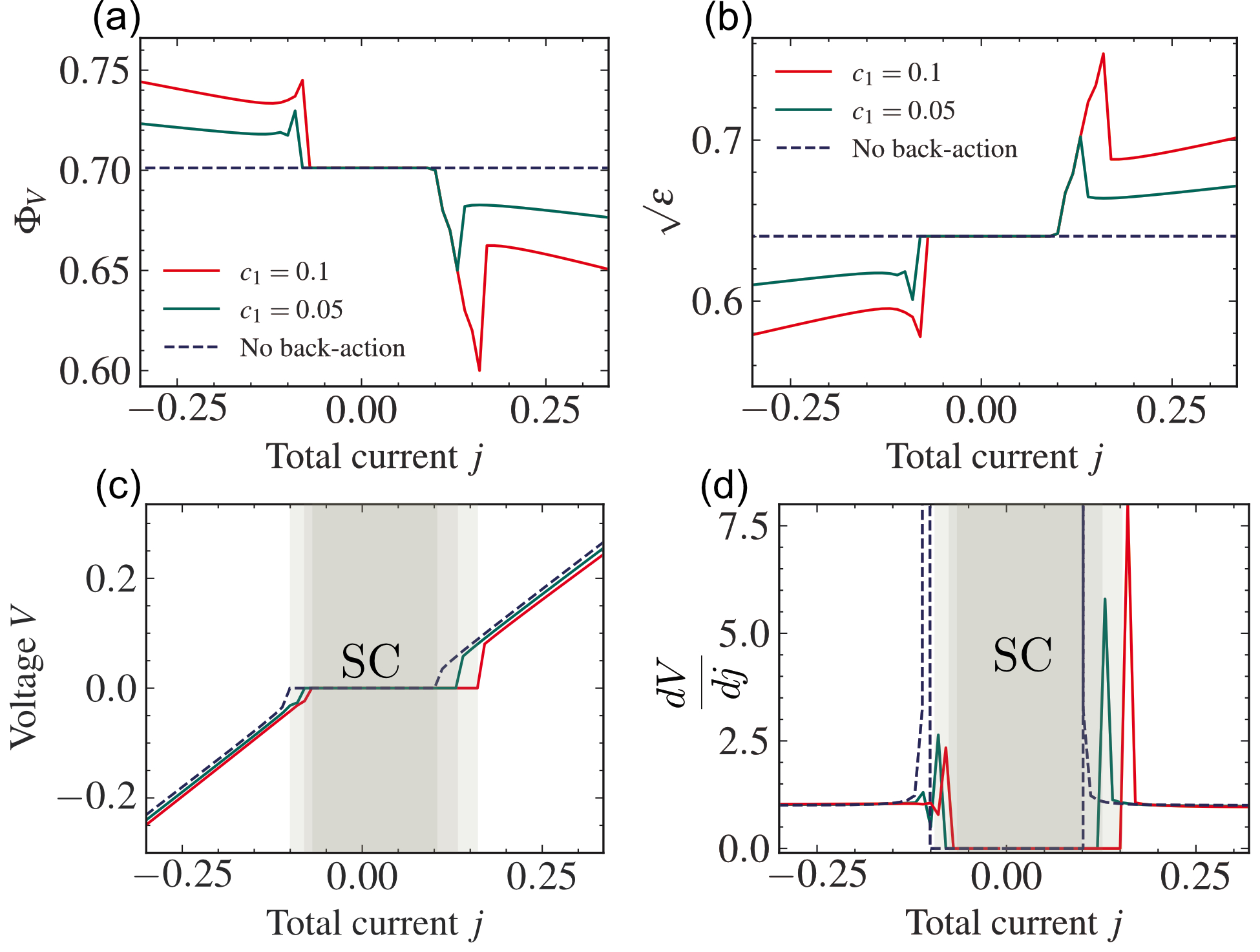}
    \caption{Regime I where $\Phi_{V}<\Phi_{V}^{c}$. (a) $\Phi_{v}$ and (b) $\sqrt{\epsilon}$ as a function of the total current $j$; (c) shows the current-voltage characteristics, and (d) the differential resistance ($\frac{dV}{dj}$) also as a function of $j$. We fix $\Phi_{V}^{0} = 0.7$ and the solid green and red curves correspond to two non-zero values of the back-action parameter $c_{1}$. The dashed lines refer to $c_1=0$ (no back action).}
    \label{fig:VI_characteristics_regime1}
\end{figure}

\begin{figure}[tb]
    \centering
\includegraphics[width=\linewidth]{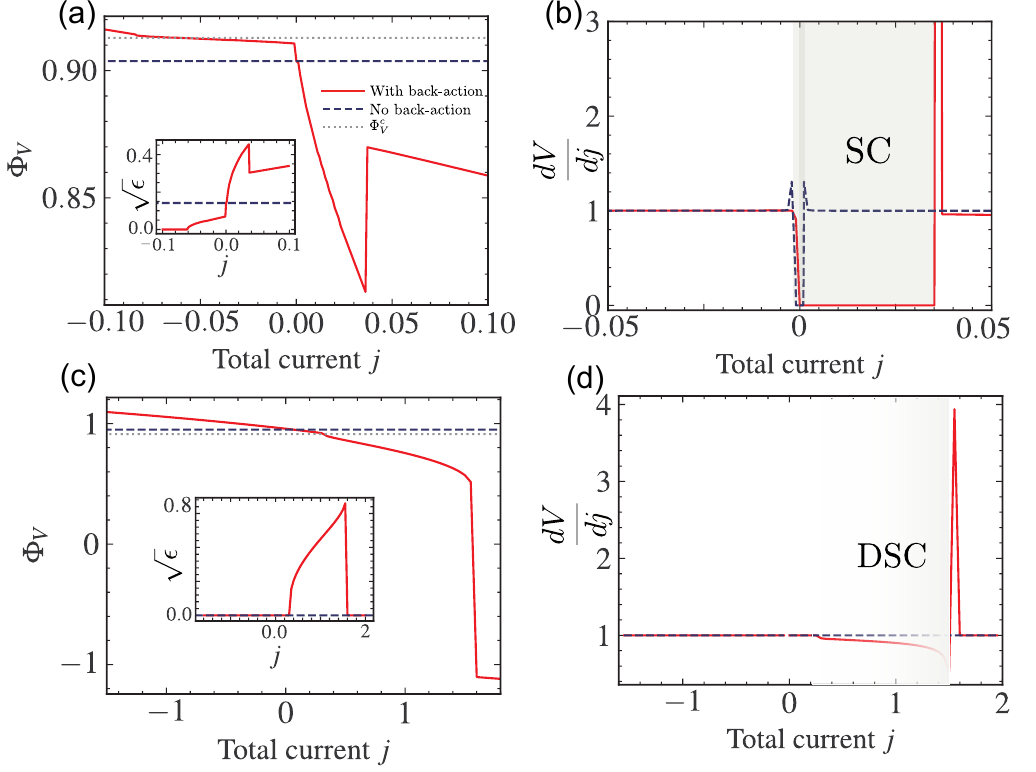}
    \caption{Effect of back action in Regimes II (a,b) and III (c,d). The first row corresponds to Regime II where $\Phi_{V}<\Phi_{V}^{c}$ showing (a) $\Phi_{V}, \sqrt{\epsilon}$ and (b) differential resistance $\frac{dV}{dj}$, respectively, as a function of the total current $j$; panels (c) and (d) shows the same but for Regime III where $\Phi_{V}>\Phi_{V}^{c}$.  We choose $\Phi_{V}^{0} = 0.90$ for Regime II and $\Phi_{V}^{0} = 0.95$ for Regime III while using the same $c_{1} = 0.25$. The  solid red lines and dashed blue lines denote the cases with and without back action respectively. In case of Regime III, the dissipative superconductor (DSC) is shown in cascading gray.}
    \label{fig:VI_characteristics_regime2}
\end{figure}

\textit{Regime I:  Away from the critical VP.---}To begin, we consider an intermediate value of VP away from the critical value, $\Phi_{V}<\Phi_{V}^{c}$. For a given total current $j$, we self-consistently solve \equsref{eq:VI_characteristics}{eq:free_energy}. 
As shown in \figref{fig:VI_characteristics_regime1}, the current-independent VP we obtain for $c_1=0$ (dashed line) results in zero SDE. However, when the coupling $c_{1}$ is switched on, the behavior changes significantly. For positive currents, $\Phi_{V}$ remains constant up to a threshold beyond which it rapidly decreases with increasing currents, through the back-action mechanism since $j_{N} \neq 0$. Consequently, the superconductivity strength characterized by $\sqrt{\epsilon}$ increases with positive currents as is seen in the second panel [see \figref{fig:VI_characteristics_regime1}(b)]. 
This enhanced superconductivity shifts the positive critical current to higher values compared to the case without back action. Beyond this region, further increases in applied current start to disrupt superconductivity, stabilizing $\Phi_{V}$ and we end up in a normal state, as is clearly seen in \figref{fig:VI_characteristics_regime1}(c,d), which shows the voltage $V$ and differential resistance $dV/dj$ against the total current $j$. 

For negative currents, the situation is quite the opposite. Here, the back-action mechanism enhances VP, thereby destabilizing superconductivity, resulting in an overall reduction of the negative critical current. This distinct behavior for positive and negative $j$ leads to a non-vanishing SDE. Increasing the back-action parameter $c_{1}$ amplifies this non-reciprocity as is evident from the green ($c_{1} = 0.05$) and red curves ($c_{1} = 0.1$). We emphasize that this SDE is solely a consequence of the dissipative back-action mechanism.

\textit{Regime II: Just below the critical VP.---}Now we consider the case $\Phi_{V}\lesssim \Phi_{V}^{c}$ where the VP at zero current is very close to but below the critical value. In this case, the critical currents needed to destroy superconductivity without back action are much smaller than in the previous regime. As we turn on $c_{1}$, we observe in \figref{fig:VI_characteristics_regime2}(a) that $\Phi_{V}$ falls sufficiently from its value at $j=0$ and $\sqrt{\epsilon}$ (shown in the inset) increases significantly compared to that without the back action on the side with positive currents. This stabilizes superconductivity even at currents $j>j_{c}|_{c_{1}=0}$, thus leading to a higher critical current. At the same time, superconductivity is destroyed much faster on the opposite side with negative currents, as the current coupling on the negative side increases the VP above the critical value very quickly, leading to a normal state. In \figref{fig:VI_characteristics_regime2}(b), we see the highly skewed non-reciprocal differential resistance (in red) as a function of current as a consequence of back action. A subtle feature is observed for higher positive currents ($j>j_{c}$) where the differential resistance starts decreasing from the value 1. This is feature occurs chiefly as a consequence of the competition between increasing total current and decreasing VP in destroying superconductivity. While for intermediate values of current ($j \gtrsim j_{c}$), we have a normal state where the current dominates the decrease in VP, there comes a point where the decrease in VP is so much that superconductivity starts getting stabilized again even at higher currents. Postponing further discussions of this feature in the next regime, we would like to remark and point out the extremely large current asymmetry observed for $\Phi_{V}\lesssim \Phi_{V}^{c}$. If we use the efficiency parameter $\eta$ to quantify the degree of SDE, the efficiency will be around $\eta \sim 96 \%$ i.e $j_{c}^{+} \sim 50 j_{c}^{-}$. All of this seems to indicate that this is probably the regime in which the extreme SDE is observed in the experiment \cite{lin_zero-field_2022}.

\textit{Regime III:  Just above the critical VP.---}We now transition on to the most exotic regime, where $\Phi_{V}\gtrsim\Phi_{V}^{c}$. Naturally, without the back-action mechanism ($c_{1}=0 $), the system remains in a normal state throughout the range of currents, as the VP exceeding the critical value only supports Ohmic behavior. Remarkably, as we switch on the back action, a positive current applied to the system can drive the system away from a completely normal state into a highly dissipative superconducting state even though at zero current the system was in a normal state, see  \figref{fig:VI_characteristics_regime2}(c,d). 
We call this a highly dissipative state, as it exhibits significant contributions of both normal and supercurrents leading to $0<\frac{dV}{dj}<1$. 

We emphasize that this is an entirely current-induced and, hence, non-equilibrium phenomenon that can be attributed to the back-action mechanism where the positive currents coupling to VP decreases it to such an extent that $\Phi_{V} <\Phi_{V}^{c}$; this enables dissipative superconductivity to become stabilized by an applied current, as seen in \figref{fig:VI_characteristics_regime2}(c). At $j \sim 0.3$, the fraction of the total current carried by the supercurrent starts increasing and leads to a fall in the differential resistance marking a departure from the Ohmic behavior as can be seen very clearly from the differential resistance characteristics in \figref{fig:VI_characteristics_regime2}(d). This deviation occurs up to a point until the applied current flips the sign of the VP (see \cite{balents_current,our_backaction}) at which point the system returns to the normal state with Ohm's law-like behavior. On the side with negative currents, this normal state is quickly reached, maintaining Ohmic behavior throughout.

\textit{Conclusion and outlook.---}In summary, we have presented a mechanism for the superconducting diode effect which is based on the coupling of the dissipative part of the current to a time-reversal-symmetry breaking order parameter, such as valley polarization relevant to the case of twisted \cite{lin_zero-field_2022,scammell_theory_2022} and rhombohedral \cite{rhombohedral} multilayer graphene. We found that it can give rise to a large current asymmetry, with the critical current asymptotically vanishing in one direction while staying finite in the other. On a more general level, our work demonstrates how non-equilibrium effects can play a crucial role for the diode effect and that studying the interplay of superconductivity and symmetry-breaking order becomes very rich when a current is applied.

\begin{acknowledgments}
M.S.S. thanks H.~D.~Scammell and J.~Li for previous collaborations on the superconducting diode effect in graphene systems. S.B acknowledges discussions with A. Rastogi, J. Sobral and L. Pupim.
All authors further acknowledge funding by the European Union (ERC-2021-STG, Project 101040651---SuperCorr). Views and opinions expressed are however those of the authors only and do not necessarily reflect those of the European Union or the European Research Council Executive Agency. Neither the European Union nor the granting authority can be held responsible for them.

\end{acknowledgments}

\bibliography{draft_Refs}

\onecolumngrid

\begin{appendix}

\section{Resistive state in superconductor: I-V characteristics}\label{appendix:resistive_state}
In this part of the appendix, we delve into the phenomenon of resistive states in superconductors in more detail. We begin by considering the full-time-dependent generalized Ginzburg-Landau equation Eq.~\eqref{full_tdgl} in one dimension without the presence of magnetic field. Substituting the ansatz $\psi =  \Delta e^{i\phi}$ (we choose a gauge where $\Delta \in \rm I\!R^{+}$) in the dimensionless equation with $\vec{A} = 0$ and separating the real and the imaginary parts of the complex equation yields two real coupled equations of $\Delta$ and phase $\phi$ as,
\begin{equation}
\Gamma_{\Delta} \frac{\partial\Delta}{\partial t}-\left(\epsilon-\Delta^{2}-\left(\frac{\partial \phi}{\partial x}\right)^2\right)\Delta-\frac{\partial^{2}\Delta}{\partial x^2}=0 .
\end{equation}
and 
\begin{equation}
\Gamma_{\Phi}\left(\mu+\frac{\partial \phi}{\partial t}\right)\Delta^2-\partial_x\left(\Delta^2 \frac{\partial \phi}{\partial x}\right)=0.
\end{equation}
where $\Gamma_{\Delta} = u \sqrt{1+\gamma^{2}\Delta^{2}}$ and $\Gamma_{\Phi} = u(\sqrt{1+\gamma^{2}\Delta^{2}})^{-1}$. Given $\vec{A}=0$, we can can introduce new gauge-invariant potentials as $\Phi = \mu + \frac{\partial \phi}{\partial t} $ and $Q = -\partial_{x} \phi$, which  we can rewrite the complete equations for $\Delta$ and $\Phi$ as, 
\begin{equation}
\Gamma_{\Delta} \frac{\partial\Delta}{\partial t}-\left(\epsilon-\Delta^{2}-Q^2\right)\Delta-\frac{\partial^{2}\Delta}{\partial x^2}=0 .
\label{eq:tdgl_delta}
\end{equation}
and 
\begin{equation}
\Gamma_{\Phi}\Phi\Delta^2+\partial_x\left(\Delta^2 Q\right)=0.
\label{eq:tdgl_phi}
\end{equation}
This is supplemented by the expression of current,
\begin{equation}
    j =-\frac{\partial Q}{\partial t}-\partial_{x} \Phi+j_{s}, \quad j_{s} =-\Delta^2 Q. 
    \label{eq:tdgl_current}
\end{equation}
Following \cite{resistive_ref}, distances are measured in units of coherence length $\xi(T)$ and $\Delta$ in units of equilibrium order parameter $\Delta_{\text{GL}}$. Additionally, one can define an electric field penetration depth $l_{E} \propto \gamma^{1/2}$.

\begin{figure}
    \centering
\includegraphics[width=0.5\textwidth]{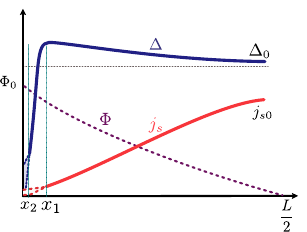}
    \caption{Schematic figure of $\Delta$, $j_{s}$ and $\Phi$ close to a phase slip. The region $0<x<x_{2}$ is the region where both $\Delta$ and $j_{s}$ are oscillatory whereas in the region $x_{2}- x_{1}$ only $j_{s}$ is oscillatory as a function of time. We are interested in the region beyond $x_{1}$ where the quantities $j_{s},\Delta$ are non-oscillatory. The potential $\Phi$ is maximum at the region of the phase slip given by $\Phi_{0}$ and decreases gradually as one goes away from the phase slip center.}
    \label{fig:phase_slip_schematic}
\end{figure}

To further quantify this phenomenon, we consider the case of $\gamma \gg 1$ which corresponds to the experimentally relevant case where the (energy spectrum of the) superconductor has a gap, and corresponds to the case where the electric field penetration depth is larger compared to the coherence length i.e. $l_{E} \propto  \gamma^{1/2} \gg 1$ \cite{resistive_ref}.We can then separate the physics into three different regions (see \figref{fig:phase_slip_schematic}):
(i) $x < x_{2}$ which corresponds to very short distances from the PSC itself ($x_{2} \sim \gamma^{-1/2} \ll 1 $) where both $\Delta$ and $\Phi$ are dynamical and strongly oscillatory; (ii) $x_{2}<x<x_{1}$  ($x_{2}<x_{1}\sim  \gamma^{1/4}$) where $j$ is oscillatory but $\Delta$ is almost non-oscillatory and
(iii) at distances far away from a PSC ($x>x_{1}$). The midpoint between two PSCs is at $x= \pm L/2$ where $\Delta = \Delta_{0}$ and $j_{s} = j_{s0}$. 

Even though the system is inherently dynamical since the static region where the potential is formed ($\sim l_{E}$) is significantly larger than the strongly dynamical region near a PSC, we can effectively consider a static model to determine our current-voltage properties \cite{resistive_ref}. 
We begin by considering region (iii) where 
the variables $\Delta, Q, \Phi$ are effectively constant in time leading us to utilise the equations \eqref{eq:tdgl_delta}, \eqref{eq:tdgl_phi}, \eqref{eq:tdgl_current}in their time-averaged form
\begin{subequations}
\begin{equation}
    \frac{\partial^2 \Delta}{\partial x^2}+\left(\epsilon-\Delta^2-\frac{j_{\mathrm{s}}^2}{\Delta^4}\right) \Delta =0
    \label{eq:approxtdgl_delta}
\end{equation}
\begin{equation}
    j =-\frac{\partial \Phi}{\partial x}+j_{\mathrm{s}}
    \label{eq:approxtdgl_j}
\end{equation}
\begin{equation}
 \Delta  \frac{u}{\gamma} \Phi=\frac{\partial j_{\mathrm{s}}}{\partial x}  
  \label{eq:approxtdgl_phi}
\end{equation}
\end{subequations}
For $x > x_{1}$, the term $\partial^2 \Delta / \partial x^2$ can be neglected and therefore,
\begin{equation}
    j_{\mathrm{s}}=\Delta^2 \sqrt{\left(\epsilon-\Delta^2\right)}
\end{equation}
We can now multiply Eq. \eqref{eq:approxtdgl_j} with Eq. \eqref{eq:approxtdgl_phi} and then integrate from $x>x_{1}$ to $x= L/2$ on both sides as, 
\begin{equation}
    \Phi^2=\frac{2\gamma}{u } \int_{j_{c}}^{j_{s_0}} \frac{\left(j-j_{\mathrm{s}}\right) d j_{\mathrm{s}}}{\Delta\left(j_{\mathrm{s}}\right)}=\frac{2\gamma}{u} \int_{\Delta_0}^{\Delta_{c}} \frac{\left[j-\Delta^2 \sqrt{\left(\epsilon-\Delta^2\right)}\right]\left(3 \Delta^2-2\epsilon\right) d \Delta}{\sqrt{\left(\epsilon-\Delta^2\right)}},
    \label{eq:int1}
\end{equation}
where $j_{s_0}$ and $\Delta_{0}$ is the supercurrent and the order parameter at $x=L/2$ and $j_{c}, \Delta_{c}$ are the supercurrent and order parameter at $x$. We also used $\Phi(L/2) = 0 $.  
Similarly, we can also integrate Eq. \eqref{eq:approxtdgl_phi} and we obtain,

\begin{equation}
    \frac{L}{2}-x=\frac{\gamma}{u } \int_{j_c}^{j_{s_0}} \frac{d j_s}{\Delta\left(j_s\right) \Phi\left(j_s\right)}=\frac{\gamma}{u } \int_{\Delta_0}^{\Delta_{c}} \frac{\left(3 \Delta^2-2\epsilon\right) d \Delta}{\sqrt{\left(\epsilon-\Delta^2\right)} \Phi(\Delta)}.
    \label{eq:int2}
\end{equation}

Before proceeding further, we also look at region (ii) $x_{2}<x<x_{1}$ where $Q,j$ are oscillatory but $\Delta$ is non-oscillatory with a position dependence as 
\begin{equation}
   \Delta=\sqrt{\epsilon}\tanh (\frac{x}{ \sqrt{2}}\sqrt{\epsilon}).
\end{equation}
In this region, $\Delta$ reaches the value of $\sqrt{\epsilon}$ at $x > 1$. We shall henceforth denote the point where $\Delta$ reaches $\sqrt{\epsilon}$ as $x_{0}$.
Additionally by combining Eq. \eqref{eq:tdgl_current} and \eqref{eq:tdgl_phi} and using $\partial_{x} j =0$, we can write the following equation
  \begin{equation}
      \frac{\partial^2 \Phi}{\partial x^2}+\frac{\partial^2 Q}{\partial x \partial t}=\Delta^2 \Gamma_{\Phi} \Phi
  \end{equation}
  If we now take the time average of this equation and use the fact that $\gamma \gg 1$, we obtain, 
  \begin{equation}
         \frac{\partial^2 \Phi}{\partial x^2} = \frac{u}{\gamma} \Delta \Phi, \quad \implies \Phi = c_{1}e^{\sqrt{u \Delta/\gamma}x} + c_{2}e^{-\sqrt{u \Delta/\gamma}x}
  \end{equation}
  We can thus conclude that in the region $x_{2} \ll x<x_{1}$ ($x \ll l_{E}$) for $\gamma \gg 1$ the time-averaged $\Phi$ does not vary since $1/\sqrt{u\Delta/\gamma} \gg \delta x =x_{1}-x_{2}$ and can therefore be represented as $\Phi \sim \Phi_{0}$.
  
The expression for electric field is given as  $E=-\partial Q / \partial t-\partial \Phi / \partial x$. If we now perform a time and space average we get,
\begin{equation}
    V = 2\frac{\Phi_{0}}{L}
    \label{eq:electric_field}
\end{equation}
where $V$ denotes the time and space average over $E$.
As noted earlier at $x_{0}$, $\Delta$ reaches its maximum ($\Delta =\sqrt{\epsilon}$) before saturating to a lower $\Delta = \Delta_{0}$ as the supercurrent increases and destabilises $\Delta$.   
We can now use the Eqs. \eqref{eq:int1} and \eqref{eq:int2}, with the upper boundary condition as $\Delta = \sqrt{\epsilon}, \Phi = \Phi_{0}$ at $x_{0}$. This leads us to the following equations,
\begin{equation}
      \Phi_{0}^2=\frac{2\gamma}{u} \int_{\Delta_0}^{\sqrt{\epsilon}} \frac{\left[j-\Delta^2 \sqrt{\left(\epsilon-\Delta^2\right)}\right]\left(3 \Delta^2-2\epsilon\right) d \Delta}{\sqrt{\left(\epsilon-\Delta^2\right)}},
\end{equation}
\begin{equation}
    L-2x_{0}=\frac{2\gamma}{u } \int_{\Delta_0}^{\sqrt{\epsilon}} \frac{\left(3 \Delta^2-2\epsilon\right) d \Delta}{\sqrt{ \left(\epsilon-\Delta^2\right)} \Phi(\Delta)}.
\end{equation}
However since $L \gg x_{0}$, we can neglect $x_{0}$ on the RHS of the equation and arrive at
\begin{equation}
    L=\frac{2\gamma}{u } \int_{\Delta_0}^{\sqrt{\epsilon}} \frac{\left(3 \Delta^2-2\epsilon\right) d \Delta}{\sqrt{ \left(\epsilon-\Delta^2\right)} \Phi(\Delta)}.
\end{equation}
Substituting this into Eq.\eqref{eq:electric_field} we get,
\begin{equation}
    V \int_{\Delta_0}^{\sqrt{\epsilon}} \frac{\left(3 \Delta^2-2\epsilon\right) d \Delta}{\sqrt{\left(\epsilon-\Delta^2\right)} F_{\Phi}\left(\Delta, \Delta_0\right)}=2 F_{\Phi}\left(\Delta=\sqrt{\epsilon}, \Delta_0\right)
\end{equation}
\begin{equation}
    F_{\Phi}^2\left(\Delta, \Delta_0\right)=\int_{\Delta_0}^{\Delta} \frac{\left[j-x^2\sqrt{\left(\epsilon-x^2\right)}\right]\left(3 x^2-2\epsilon\right) d x}{\sqrt{\left(\epsilon-x^2\right)}} .
\end{equation}
These are the equations that we study in the main text. While we have considered the case with $j>0$, a similar analysis can be conducted for $j<0$. To further motivate the choice of $j_{s0}, \Delta_{0}$, we consider the equation $ j_{\mathrm{s}}=\Delta^2 \sqrt{\left(\epsilon-\Delta^2\right)}$ and consider the maximum possible supercurrent in between the PSCs. This fixes the lower bound as $ \Delta_{0} =  \sqrt{2\epsilon/3}$ corresponding to a critical supercurrent of  $j_{s0} = 2 \epsilon^{3/2}/3\sqrt{3}$.  
 We also note that if the total current is less than $j_{s0}$ the voltage will be zero as it is a dissipationless superconductor with no phase slips inducing voltage drops. Beyond $j>j_{s0}$, we enter the resistive state and solve the equations above numerically by calculating the voltage as a function of the total current $j$.

\section{Particle-particle bubble expression in the presence of valley polarisation}\label{appendix:particlebubble}
Here we look at the expression for the particle-particle bubble in the presence of the TRS breaking normal state order -- valley polarisation. We begin by stating the electronic Hamiltonian, 
\begin{align}\begin{split}
    \mathcal{H}_c = &\sum_{\vec{k},\nu} \xi_{\vec{\vec{k}},\nu} c^\dagger_{\vec{k},\nu}c^\pdagger_{\vec{k},\nu} +  \Phi_V \sum_{\vec{k},\nu} \nu c^\dagger_{\vec{k},\nu}c^\pdagger_{\vec{k},\nu}+ \sum_{\vec{k},\vec{q}} \left[ \Delta_{\vec{q}} c^\dagger_{\vec{k}+\vec{q}/2,+}c^\dagger_{-\vec{k}+\vec{q}/2,-}  + \text{H.c.} \right], \label{CouplingToVPandSC}
\end{split}\end{align}
where $c_{\vec{k},\nu}^\dagger$ creates an electron in valley $\nu=\pm$ and at momentum $\vec{k}$, $\xi_{\vec{k},\nu}$ encodes the non-interacting band-structure, and the valley imbalance is denoted by $\Phi_{V}$. We can then perform a Hubbard-Stratonovich transformation in the intervalley Cooper channel for a given $\Phi_{V}$ to obtain $\mathcal{H}_{\text{HS}} = \mathcal{H}_{c} +\sum_{\vec{q}}\frac{|\Delta_{\vec{q}}|^{2}}{g}$. Integrating out the electrons, the associated expression for the change in free energy $\delta \mathcal{F}_{\text{S}} = \mathcal{F}_{\text{S}}(\Delta_{\vec{q}},\Phi_{V}) - \mathcal{F}_{\text{S}}(0,\Phi_{V})$ for superconductivity is given by
\begin{equation}
 \delta \mathcal{F}_{\text{S}} \sim \sum_{\vec{q}} a^{\text{S}}_{\vec{q}} \, |\Delta_{\vec{q}}|^2 + \mathcal{O}(|\Delta_{\vec{q}}|^4),
\end{equation}
where the coefficient $a^{\text{S}}_{\vec{q}} = \frac{1}{g} - \Gamma_{\vec{q}}$ is evaluated microscopically from the above Hamiltonian to be of the form
\begin{equation}
    \Gamma_{\vec{q}} = \frac{1}{2N}\sum_{\vec{k}} \frac{\tanh{\frac{E_{\vec{k},\vec{q},+}}{2 T}} + \tanh{\frac{E_{\vec{k},\vec{q},-}}{2 T}}}{E_{\vec{k},\vec{q},+} + E_{\vec{k},\vec{q},-}}. \label{ParticleParticleBubble}
\end{equation}
where $E_{\vec{k},\vec{q},\nu} =  \xi_{\vec{k} + \nu \vec{q}/2} + \nu \,\Phi_V$.
If we now expand $a^{\text{S}}_{\vec{q}}$ in orders of $\Phi_{V}$, we obtain
\begin{equation}
    a^{\text{S}}_{\vec{q}} \sim \frac{1}{g} -  \alpha_{\vec{q}} +\alpha'_{\vec{q}} \Phi_{V}  - \alpha''_{\vec{q}} \Phi_{V}^{2} + \mathcal{O}(\Phi_{V}^{4})
\end{equation}
where 
\begin{subequations}
    \begin{equation}
        \alpha_{\vec{q}}  = \frac{1}{2N}\sum_{\vec{k}} \frac{\tanh{\frac{ \xi_{\vec{k} +\vec{q}/2}}{2 T}} + \tanh{\frac{ \xi_{\vec{k} -  \vec{q}/2}}{2 T}}}{ \xi_{\vec{k} +  \vec{q}/2} +  \xi_{\vec{k} - \vec{q}/2}}, \quad  \alpha'_{\vec{q}}  = \frac{1}{4N T}\sum_{\vec{k}} \frac{\tanh^{2}{\frac{ \xi_{\vec{k} +\vec{q}/2}}{2 T}} - \tanh^{2}{\frac{ \xi_{\vec{k} -  \vec{q}/2}}{2 T}}}{ \xi_{\vec{k} +  \vec{q}/2} +  \xi_{\vec{k} - \vec{q}/2}}  
    \end{equation}
    \begin{equation}
        \alpha''_{\vec{q}}  = \frac{1}{8N T^{2}}\sum_{\vec{k}} \frac{-\tanh{\frac{\xi_{\vec{k} +\vec{q}/2}}{2 T}} + \tanh^{3}{\frac{ \xi_{\vec{k} +\vec{q}/2}}{2 T}} -\tanh{\frac{ \xi_{\vec{k} -  \vec{q}/2}}{2 T}}+ \tanh^{3}{\frac{ \xi_{\vec{k} -  \vec{q}/2}}{2 T}}}{ \xi_{\vec{k} +  \vec{q}/2} +  \xi_{\vec{k} - \vec{q}/2}}
    \end{equation}
\end{subequations}
In our case, TRS/inversion implies $\xi_{\vec{k}}=\xi_{\vec{-k}}$, which leads to $\alpha^{\prime} = 0$. Therefore the leading order contribution to $a_{\vec{q}}$ comes from the term quadratic in $\Phi_{V}$. This analysis motivates our choice of $\epsilon$ in the main text which we take to depend on $\Phi_{V}$ quadratically. 

\end{appendix}

\end{document}